\documentclass[sigconf]{acmart}
\usepackage{amsmath,nccmath,tabularx} %amssymb
\usepackage{algorithm}
\usepackage{algorithmic}
\usepackage{color}
\usepackage{graphicx}
\graphicspath{{images/}}
\usepackage{wrapfig,lipsum}
\usepackage{tabularx}
\usepackage{graphicx}
\usepackage{lscape}
\usepackage{subcaption}
\usepackage{latexsym}
\usepackage{pifont}% http://ctan.org/pkg/pifont
\usepackage{multirow}

\usepackage{enumitem}
\setlist{leftmargin=6mm}
\usepackage[T1]{fontenc}
\usepackage[utf8]{inputenc}
\usepackage[font=small,labelfont=bf]{caption}

\usepackage{booktabs}

\usepackage{xcolor}
\usepackage{xspace}
\usepackage{bm}
\usepackage{amsmath}
\usepackage{placeins}

\usepackage{subcaption}
\usepackage{listings}
\usepackage{balance} 

\AtBeginDocument{%
  \providecommand\BibTeX{{%
    \normalfont B\kern-0.5em{\scshape i\kern-0.25em b}\kern-0.8em\TeX}}}

\begin{document}
%\fancyhead{}
%\title{Is Vec2Text Really a Threat to Dense Retrieval Systems?}

%ORIGINAL (ARXIV):
\title{Understanding and Mitigating the Threat of Vec2Text to Dense Retrieval Systems}

%Alternatives:
%Pick-the-Best Prompt: Streamlining Efficient Text Ranking
%Top-Pick Prompt, Champion-Selection Prompt


\author{Shengyao Zhuang}
\affiliation{%
	\institution{CSIRO}
	%\streetaddress{4072 St Lucia}
	\city{Brisbane}
	%\state{QLD}
	\country{Australia}}
\email{shengyao.zhuang@csiro.au}
\author{Bevan Koopman}
\affiliation{%
	\institution{CSIRO, The University of Queesnland}
	%\streetaddress{4072 St Lucia}
	\city{Brisbane}
	%\state{QLD}
	\country{Australia}}
\email{bevan.koopman@csiro.au}

\author{Xiaoran Chu}
\affiliation{%
	\institution{The University of Queensland}
	%\streetaddress{4072 St Lucia}
	\city{Brisbane}
	%\state{QLD}
	\country{Australia}}
\email{xiaoran.chu@uq.edu.au}

\author{Guido Zuccon}
\affiliation{%
	\institution{The University of Queensland}
	%\streetaddress{4072 St Lucia}
	\city{Brisbane}
	%\state{QLD}
	\country{Australia}}
\email{g.zuccon@uq.edu.au}

%%\renewcommand{\shortauthors}{Trovato and Tobin, et al.}

\begin{abstract}
	
The emergence of Vec2Text --- a method for text embedding inversion --- has raised serious privacy concerns for dense retrieval systems which use text embeddings, such as those offered by OpenAI and Cohere. This threat comes from the ability for a malicious attacker with access to  embeddings to  reconstruct the original text.

In this paper, we investigate various factors related to embedding models that may impact text recoverability via Vec2Text. We explore factors such as distance metrics, pooling functions, bottleneck pre-training, training with noise addition, embedding quantization, and embedding dimensions, which were not considered in the original Vec2Text paper. Through a comprehensive analysis of these factors, our objective is to gain a deeper understanding of the key elements that affect the trade-offs between the text recoverability and retrieval effectiveness of dense retrieval systems, offering insights for practitioners designing privacy-aware dense retrieval systems. We also propose a simple embedding transformation fix that guarantees equal ranking effectiveness while mitigating the recoverability risk.
%Further, we also adapt Vec2Text to the seperate task of corpus poisoning, where theoretically, Vec2Text is far more dangerous than previous attack methods (Vec2Text does not need access to the dense retriever's model parameters and it can efficiently generate numerous adversarial passages). 

Overall, this study reveals that Vec2Text could pose a threat to current dense retrieval systems, but there are some effective methods to patch such systems\footnote{Code available at \url{https://github.com/ielab/vec2text-dense_retriever-threat}}.

\end{abstract}



%\ccsdesc[500]{Information systems~Information retrieval query processing}
%%
%%Keywords. The author(s) should pick words that accurately describe

\maketitle

\section{Introduction}

Text embeddings are dense vector representations which capture semantic information about the text they encode~\cite{muennighoff-etal-2023-mteb}. Search engines that leverage these embeddings often employ dense retrievers (DRs)~\cite{tonellotto2022lecture,zhao2022dense,guo2022semantic,bruch2024foundations}. These retrievers utilize text embedding models to encode both queries and documents into embeddings. A similarity metric (e.g., cosine similarity) is then used to estimate relevance. DRs have demonstrated improved retrieval effectiveness compared to traditional exact term-matching search systems, arguably due to the rich semantic information encoded in embeddings~\cite{yates-etal-2021-pretrained}.

However, a recent study conducted by~\citet{morris-etal-2023-text} raises serious privacy concerns regarding DRs. This study explored the issue of \textit{inverting} textual embeddings: recovering the original text from its embedding. The proposed Vec2Text method iteratively corrects and generates text to reconstruct the original text based on the given input embedding. According to the results presented in the original paper, Vec2Text can accurately recover 92\% of short text and reveal sensitive information (such as patient names in clinical notes) with high accuracy.

Even more concerning is that training Vec2Text does not require access to the embedding model parameters; only the text-embedding pairs from the training data are needed. This implies that DR systems that use popular text embedding API services, such as those provided by OpenAI and Cohere, can be vulnerable to attacks from malicious attackers. These attackers can build their own training data by sending numerous text snippets to the API and obtaining the corresponding embeddings. Subsequently, these text-embedding pairs can be utilized to train a Vec2Text model against the target embedding model. Once the Vec2Text model is trained, the document embeddings stored in these DR systems are no longer secure. An attacker with access to the document embeddings can recover the original text from the embeddings using the trained Vec2Text model.

A limitation of the original Vec2Text paper was that it investigated only two embedding models\footnote{GTR-base and OpenAI text-embeddings-ada-002.}. Presently, there are numerous DR systems that utilize various types of embedding models. The distinctions among these embedding models arise from different strategies employed during training or inference, such as distance metrics, pooling functions, bottleneck pre-training, embedding quantization, and embedding dimensions. These diverse strategies involve trade-offs between retrieval effectiveness and efficiency for DRs. However, the impact of these strategies on Vec2Text reconstruction remains unexplored. In this paper, we address this gap by re-training Vec2Text models against various DR systems that implement these different strategies.
Our comprehensive experimental results reveal that some strategies can make DRs more susceptible to Vec2Text, while others effectively protect privacy without compromising retrieval effectiveness. Additionally, we propose a simple yet effective embedding transformation method, secret to each user, which ensures that retrieval effectiveness remains intact while completely mitigating the risk of text reconstruction. Our study provides valuable insights for practitioners designing privacy-aware dense retrieval systems.

\begin{figure*}
	
	\includegraphics[width=1\textwidth]{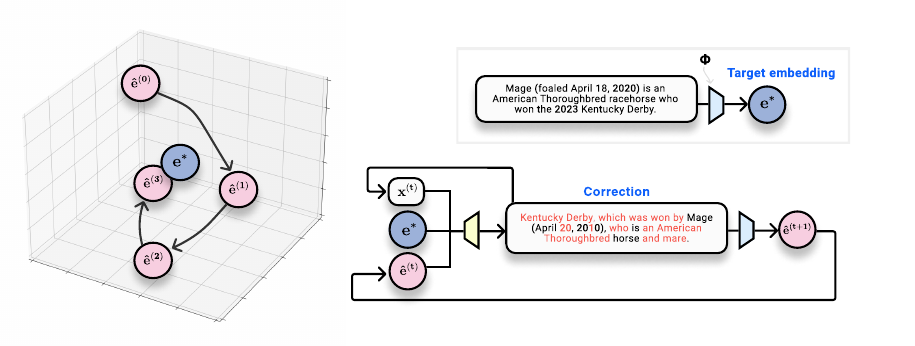}
	\caption{Overview of Vec2Text, taken from \citet{morris-etal-2023-text} with permission. ``Given access to a target embedding $e$ (blue) and query access to an embedding model $\phi$ (blue model), the system aims to iteratively generate (yellow model) hypotheses $\hat{e}$ (pink) to reach the target. Example input is a \textit{[passage]} taken from a recent Wikipedia article (June 2023). Vec2Text perfectly recovers this text from its embedding after 4 rounds of correction.''}
	\label{fig:vec2text}
\end{figure*}

Our contributions can be summarized in four main aspects:

\begin{enumerate}
	\item We reproduce Vec2Text, correcting an embedding model implementation error and a testing data leakage error in the original experimental setting. This provides a more accurate evaluation of Vec2Text reconstructibility effectiveness and its relation to the embedding model retrieval effectiveness.
	%We reproduce Vec2Text, fixing a serious experimental error in the use of the output layer of the GTR-base embeddings. This fix has implications in terms of both retrieval and reconstructibility effectiveness, although the trends observed in the original work do not change.  %and thus improving effectiveness.
	\item We re-train Vec2Text against a wide range of embedding models to understand the key elements that contribute to the text reconstruction.
	\item Our comprehensive results provide insights into the trade-offs between privacy protection and retrieval effectiveness.
	\item We propose an embedding transformation method to mitigate privacy risks without compromising retrieval performance.

\end{enumerate}

%\section{Related Works?}
%contextualise vec2text with respect to other works. Also for corpus poisoning.
%
%Maybe move to end of paper?

\section{The Vec2Text Method}

Vec2Text inverts the text encoding process: given an input embedding, generate the text it represents~\cite{morris-etal-2023-text}. (A graphical overview of Vec2Text, taken from \citet{morris-etal-2023-text} with permission, is shown in Figure~\ref{fig:vec2text}.) The method comprises two training stages. 
In the initial stage, a hypothesis text generation model is trained, utilizing a conditional transformer generative model that exclusively takes the embedding as the model input. The training objective is to produce the original text. However, \citeauthor{morris-etal-2023-text} found that this simplistic model is insufficient for generating highly accurate original text. Consequently, they consider the text generated by this first stage model as a hypothesis.

Moving on to the second stage, the learning process involves training another transformer generative model. This model aims to generate satisfactory text by refining the initial hypothesis generated in the first stage. The refinement is achieved through an iterative feedback, re-embedding and correction process. In each iteration step, the model takes the ground-truth  embedding, the generated text, and its embedding from the last iteration step as inputs (the 0 step uses the hypothesis text generated from the first stage model). The output target is the original text. This iterative process allows the model to focus on the differences between the generated text and the original text in the embedding space and gradually reduce these differences.

The models in Vec2Text are parameterized as a standard encoder-decoder transformer conditioned on the previous output. One challenge is inputting conditioning embeddings into the transformer encoder, which requires a sequence of embeddings as input. To address this, a small multi layer perceptron is used to project a single embedding vector to a larger size and reshape it to match the input requirements of the encoder.

During inference, a sequence-level beam search is used to guide the generation. At each step of correction, a number of possible corrections are considered, and the top unique continuations are selected based on their distance in embedding space to the ground-truth embedding.

 \citet{morris-etal-2023-text} trained Vec2Text to invert\footnote{i.e. to reconstruct the original text that corresponds to the embeddings.} GTR-base~\cite{ni-etal-2022-large} and OpenAI's text-embeddings-ada-002 embedding models. It achieved high reconstructibility\footnote{i.e. the capability to reconstruct the original text from the embedding.} on in-domain datasets and adapted well to different-length inputs. The method outperformed baselines on various metrics, and running beam search improved the exact match score.
In a case study on MIMIC III clinical notes, Vec2Text demonstrated high performance in reconstructing GTR-embedded clinical notes, re-identifying high percentages of first names, last names, and complete names: this  raises serious privacy concerns about the embeddings.

However, we note that \citeauthor{morris-etal-2023-text} only considered reconstructing GTR-base and OpenAI text-embeddings-ada-002 embedding models. Many other different embedding models exist, and each uses very different training and inference strategies. Can all embedding models be easily attacked by Vec2Text? What types of embedding models are vulnerable to Vec2Text? In the following sections, we explore a wide range of key decisions in constructing embedding models (e.g., pooling functions, bottleneck pre-training, quantization), that go well beyond the scope of the original paper.
\section{Reproduction of vec2text}

\subsection{Experimental Methodology}

\begin{table*}[h]
	\caption{Reproduction of Vec2Text's text reconstructibility results on NQ dataset. Bug fixed represents a correction to ensure Vec2Text uses the final dense pooling layer of GTR-base for output.}\label{tab:reproduce}
	\begin{tabular}{@{}lllllllll@{}}
		\toprule
		\multirow{2}{*}{\bf Model Checkpoints}&\multicolumn{4}{c}{\bf Original NQ valid set} & \multicolumn{4}{c}{\bf Filtered NQ valid set} \\
		\cmidrule(lr){2-5} \cmidrule(lr){6-9}
		&bleu &tf1 &exact &cos &bleu &tf1 &exact &cos \\\midrule
		1. GTR-base (paper~ \cite{morris-etal-2023-text}) &97.3 &99.0 &92.0 &0.99 &- &- &- &- \\\midrule
		2. GTR-base (author checkpoint) &98.2 &99.5 &94.4 &0.998 &97.3 &99.2 &92.2 &1.0\\
		3. GTR-base (reproduced) &97.0 &99.2&91.1&1.000 &96.1 &98.9&89.0&1.0\\
		4. GTR-base (bug fixed) &97.9 &99.4 &94.0 &1.000  &96.4 &99.0 &90.1 &1.0 \\ 
		\bottomrule
	\end{tabular}
\end{table*}

\begin{table}[h]
	\caption{Retrieval effectiveness of GTR-base with/without the last dense pooling layer. $*$ denote statistically significantly difference using a paired, two-tails t-test ($p < 0.01$).}\label{tab:retrieval}
		\begin{tabular}{@{}lllll@{}}
			\toprule
			GTR-base &top 10 &top 20 &top 100 &top 1000 \\\midrule
			Without pooling layer &0.634 &0.711 &0.816 &0.893 \\
			With pooling layer &0.718$^{*}$ &0.773$^{*}$ &0.855$^{*}$ &0.913$^{*}$ \\ 
			\bottomrule
		\end{tabular}
\end{table}

We begin by completing a full reproduction of Vec2Text. This is needed to ensure a fair comparison of the different embeddings models and settings with which we further study Vec2Text.

We use the official Vec2Text codebase\footnote{\url{https://github.com/jxmorris12/vec2text}} provided by the authors and adhere to the training parameters outlined in the guidance of the code repository. Specifically, we train Vec2Text on documents from the Natural Question (NQ) dataset~\cite{kwiatkowski-etal-2019-natural}, setting the max sequence length to 32. The AdamW optimizer is employed with a learning rate of $1 \times 10^{-3}$, incorporating warmup. A batch size of 512 is utilized, and all models are trained on a single H100 GPU for 50 epochs. 
One major deviation from the original setting is the number of training epochs. We note that the complete Vec2Text training pipeline is time-consuming. The authors reported in the paper that the full pipeline takes 2 days to train for 100 epochs with four A6000 GPUs. However, in our implementation using the provided code, the full pipeline takes approximately 10 days with a single H100 GPU. Given the extensive number of embedding models we plan to investigate and the limited computational resources at our disposal, we train for 50 epochs. This reduces the training time to around 5 days for each embedding model, making the process more feasible within our constraints.

In addition to the variation in the number of epochs, we have identified two significant issues in the original experimental setting.
	
\begin{enumerate}
	\item[\bf1)] \textbf{GTR-base model missing dense pooling layer:} The first issue involves the training of Vec2Text for reconstructing text from the GTR-base embeddings. 
	\citeauthor{morris-etal-2023-text} inadvertently used the last hidden layer outputs of the GTR-base model as the final text embeddings. However, it's crucial to note that GTR-base model incorporates an additional dense pooling layer to project the last hidden layer outputs, a detail overlooked by the Vec2Text training code.
	To address this issue, this section includes a re-training of Vec2Text with the correct GTR-base embeddings, and we compare the results with the original setting.
	
	\item[\textbf{2)}] \textbf{Data leakage in NQ dataset splits:} The second issue relates to a serious data leakage problem in the author-provided Vec2Text NQ training and validation dataset splits. We found that $50.7\%$ of validation data points also appear in the training dataset. Although the authors evaluated Vec2Text also on out-domain datasets and demonstrated its reconstruction effectiveness, the reported effectiveness on the NQ dataset (in-domain evaluation) reported in the original paper was likely overestimated. Hence, in our experiments, we re-evaluate Vec2Text for the NQ datasets with a corrected, non-overlapping validation set.
	
\end{enumerate}	

\noindent These two corrections ensure a more accurate evaluation of Vec2Text's performance in inverting GTR-base embeddings.

Finally, to compare how different embedding models balance retrieval effectiveness and privacy preservation (i.e., not being reconstructed by Vec2Text), we also measure the retrieval effectiveness of all considered embedding models. Following the original Vec2Text training setting, which involves training Vec2Text on documents from the NQ dataset, we adopt the \textit{top-k retrieval accuracy} evaluation setting from DPR~\cite{karpukhin-etal-2020-dense}, which measures the percentages of retrieved passages that contain a span that answers the
question (query). This evaluation is commonly employed in the literature for assessing retrieval effectiveness on the NQ dataset~\cite{gao-callan-2022-unsupervised, wang-etal-2023-simlm, qu-etal-2021-rocketqa, ma-etal-2021-simple, xiong2020approximate}. We use the official NQ test queries and Pyserini~\cite{jimmy2021pyserini} IR toolkit for this evaluation.

To evaluate reconstructibility (i.e., the ability to recover the original text from its embedding) we adopt the same measures used by \citet{morris-etal-2023-text}: 
\begin{itemize}
	\item BLEU score~\cite{Papineni2002bleu}, a metric capturing n-gram similarities between the true and reconstructed text;
	\item Token F1, representing the multi-class F1 score between the predicted and true token sets;
	\item Exact-match, indicating the percentage of reconstructed outputs perfectly matching the ground truth;
	\item Cosine similarity, measuring the similarity between the true embedding and the embedding of the reconstructed text.
\end{itemize}

We follow the default evaluation setting in the original codebase, which requires the random sampling of 1,000 passages from the full NQ validation set. We note that the full size of the author-provided NQ validation set has 849,508 data points, of which 50.7\% also appear in the training set. To resolve this data leakage problem, we further filtered out these overlapping data points to create our filtered NQ validation set, which now consists of 419,851 data points.

For the Vec2Text generation inference configuration, we use sequence-level beam search with 50 steps and a beam width of 8. The whole evaluation for a single experiment took around 4 hours on one Nvidia H100 GPU.

\subsection{Reproduction Results}

Table~\ref{tab:reproduce} shows Vec2Text reproduction results on \citeauthor{morris-etal-2023-text}'s NQ validation set and our filtered NQ validation set. For reference, in the first row of the table we list scores as appear in the original paper. We note that these scores are obtained on \citeauthor{morris-etal-2023-text}'s validation set which is affected by the data leakage.
	
We first consider evaluating the author-provided Vec2Text model checkpoint, which is supposed to be the same as the checkpoint evaluated and reported in the original paper. This experiment aims to verify the correctness of our evaluation setting and understand the impact of the data leakage issue. We report our results in the second row of Table~\ref{tab:reproduce}. Compared to the scores reported by \citeauthor{morris-etal-2023-text}, we obtained higher scores on the original validation split, likely due to the randomness of the sampled subset. The evaluation scores on our filtered validation set are only slightly lower, suggesting that the author-provided checkpoint did not overfit the training data and generalizes well to the non-overlapping data points. Thus, although there is a considerable amount of data leakage, the negative impact on the evaluation validity appears to be negligible.

In the third row, we report the results obtained from our reproduced Vec2Text model with incorrect GTR-base embeddings (i.e., trying to exactly reproduce the original Vec2Text results). We note that the major training difference from the original checkpoint is the number of training epochs; our reproduction uses half the number of epochs (50 epochs) compared to the original checkpoint (100 epochs). Despite this difference, our reproduced checkpoint only shows a slight decrease in the BLEU and exact match scores but achieves very close scores on other evaluation metrics on the original validation set, and, like the original checkpoint, only slightly lower on the filtered validation set. These results suggest that Vec2Text almost converges with half of the epochs used by \citeauthor{morris-etal-2023-text}, and the only notable improvement with longer training is in exact reconstruction. From here, we only consider Vec2Text models trained for 50 epochs, noting that their results are likely similar to those obtained if 100 epochs were used.

Finally, in the last row, we present the results using the correct GTR-base embeddings for training Vec2Text with 50 epochs. We achieve comparable Vec2Text reconstructibility with the original checkpoint, despite utilizing only half the training epochs. These results suggest that employing the correct outputs from the embedding model, rather than the intermediate layer outputs, may contribute to improved Vec2Text reconstruction. On the other hand, in Table~\ref{tab:retrieval} we report the retrieval effectiveness of the GTR-base embedding model with (correct embeddings) and without (incorrect embeddings) the last dense pooling layer. Our results emphasize the significance of using the correct embedding for retrieval, as the top-k accuracy exhibits a substantial increase compared to that obtained with incorrect embeddings.

To conclude this section, our reproduction results demonstrate that using the correct GTR-base embeddings and a reduced number of training epochs can  achieve high reconstructibility and retrieval effectiveness, highlighting the robustness and efficiency of the Vec2Text model under different configurations. Additionally, the slightly lower evaluation scores on our filtered validation set suggest that the \citeauthor{morris-etal-2023-text}'s checkpoint generalizes well and that the impact of data leakage is minimal, despite the amount of leakage itself being considerable. For the rest of the experiments in this study, we use our filtered validation set to evaluate the reconstructibility of Vec2Text models.
\begin{table*}\centering
	\caption{Reproduction of the DPR~\cite{karpukhin-etal-2020-dense} embedding models with different distance metrics and pooling functions. "cls" indicates the utilization of the CLS token embedding as the representation, while "mean" indicates the use of mean pooling, aggregating all token embeddings to generate the final text embedding. "dot" indicates the dot product distance metric and "cos" indicates the cosine similarity distance metric. There is no statistically significant difference observed in terms of retrieval effectiveness.}
	\label{tab:dpr}
%		\resizebox{1\textwidth}{!}{
		\begin{tabular}{@{}lllrrrrrrrr@{}}\toprule
		\multicolumn{3}{c}{\bf Embedding Models} & \multicolumn{4}{c}{\bf Retrieval effectiveness} & \multicolumn{4}{c}{\bf Reconstructibility} \\
		\cmidrule(lr){4-7} \cmidrule(l){8-11}
		Label & Metric & Representation
			&top 10 &top 20 &top 100 &top 1000 &bleu &tf1 &exact &cos \\\midrule
			a. DPR\_cls\_dot & Dot-product & CLS token & 0.748 &0.800 &0.863 &0.914 &79.2 &90.9 &43.0 &0.996 \\
			b. DPR\_cls\_cos & Cosine similarity & CLS token &0.748 &0.799 &0.863 &0.914 &82.3 &92.1 &50.0 &0.994\\
			c. DPR\_mean\_cos & Cosine similarity & Mean pooling &0.745 &0.802 &0.866 &0.916 &89.4 &95.2 &57.9 &0.996 \\
			\bottomrule
		\end{tabular}
%			}
\end{table*}

\section{Understanding what impacts Vec2Text effectiveness}
In the previous sections, we successfully reproduced Vec2Text with the GTR-base embedding model and finalized the experimental environment. In this section, our goal is to understand how various strategies implemented by different embedding models impact retrieval effectiveness and Vec2Text reconstructibility. To achieve this, we re-train Vec2Text with respect to the target embedding models. Each of the embedding models represents different choices in terms of distance metric, pre-training strategy, etc. When re-training, we used the same Vec2Text parameters reported above.

%against target embedding models that were trained with different strategies, using the exact same parameters.

To maintain control over experimental parameters and prevent results from being influenced by factors such as different datasets and training parameters, we independently train the embedding models ourselves. This is done with the same experimental settings but utilizing different training or inference strategies. Specifically, we start by reproducing the DPR embedding model~\cite{karpukhin-etal-2020-dense}, following the training parameters outlined in the DPR paper and employing the Tevatron dense retriever training toolkit~\cite{gao2023tevatron} to train the DPR model on the NQ dataset, the same dataset used to train Vec2Text. After reproducing the DPR retrieval results, we then apply the same dense retriever training hyperparameters (e.g., batch size, learning rate, and number of hard negatives) to train other DPR variations with distinct training strategies, as detailed in the next subsections.

\begin{table*}[h!]\centering
	\caption{Zero-shot retrieval and bottlenecked pre-training results. 
		Subscripts denote statistically significantly better retrieval effectiveness with a T-test ($p < 0.01$). Results show that even zero-shot retrievers are at risk to Vec2Text reconstruction. Bottleneck pre-training proved useful at improve retrieval effectiveness but greatly increased risk of reconstructibility.}
	\label{tab:zeroshot}
%		\resizebox{1\textwidth}{!}{
		\begin{tabular}{@{}llllllllll@{}}\toprule
			\multicolumn{2}{c}{\bf Embedding Models}&\multicolumn{4}{c}{\bf Retrieval effectiveness} & \multicolumn{4}{c}{\bf Reconstructibility} \\
			\cmidrule(lr){3-6}\cmidrule(l){7-10}
			Label & Setting& top 10 &top 20 &top 100 &top 1000 &bleu &tf1 &exact &cos \\\midrule
			a. BERT\_cls\_dot & Zero-shot & 0.101 & 0.153& 0.288 &0.467 & 47.9& 73.3& 11.7&0.981 \\
			b. BERT\_mean\_cos & Zero-shot &  0.283$^{a}$& 0.358$^{a}$& 0.520$^{a}$ & 0.715$^{a}$ & 86.9&94.1 &51.6 & 0.996\\
			c. SimLM\_cls\_dot & Bottleneck + Zero-shot  & 0.304$^{ab}$  & 0.378$^{a}$ & 0.546$^{ab}$  & 0.720$^{a}$  & 94.4& 97.2& 64.4& 0.999 \\\hline
			d. SimLM\_cls\_dot\_finetuned & Bottleneck + Fine-tuned & 0.750$^{abc}$ & 0.805$^{abc}$& 0.874$^{abc}$ & 0.919$^{abc}$ & 94.7& 97.2& 63.1& 1.000\\
			\bottomrule
		\end{tabular}
%			}
\end{table*}

\subsection{Distance Metric and Pooling Method}
In Table~\ref{tab:dpr}, we examine the impact of different distance metrics, namely dot product (dot) or cosine similarity (cos), used for training and inference in DPR models. Additionally, we investigate the pooling method employed to construct a single embedding for each input text, specifically, CLS token embedding (cls) or mean pooling of all token embeddings (mean).

The \textit{DRP\_cls\_dot} represents our reproduction of the original DPR embedding model, utilizing CLS token embedding to represent the entire text and employing the dot product during training. We achieved very similar top-k retrieval effectiveness as reported in the original DPR paper, indicating successful reproduction of DPR. It is worth noting that our trained \textit{DRP\_cls\_dot} exhibits higher retrieval effectiveness than GTR-base embedding models presented in Table~\ref{tab:retrieval}, possibly due to our DPR training and testing data originating from the same data distribution, while GTR-base has training data from other domains.

However, concerning Vec2Text reconstructibility, we observe that Vec2Text trained with the \textit{DRP\_cls\_dot} embedding model yields much lower reconstructibility compared to GTR-base embedding models. This suggests that the embeddings provided by \textit{DRP\_cls\_dot} contain less information for Vec2Text to reconstruct the original text, indicating a potentially more secure embedding model than GTR-base.

We then proceeded to train two additional DPR variations, namely \textit{DRP\_cls\_cos}, which employs CLS token embedding and cosine similarity, and \textit{DRP\_mean\_cos}, which utilizes mean pooling of all token embeddings with cosine similarity. There was no statistically significant differences in retrieval effectiveness for these settings.

Regarding reconstructibility, a comparison between \textit{DRP\_cls\_dot} and \textit{DRP\_cls\_cos} suggests that cosine similarity appears to enhance Vec2Text's scores. However, when comparing \textit{DRP\_cls\_cos} and \textit{DRP\_mean\_cos}, the reconstructibility of Vec2Text significantly increases. This surprising result suggests that mean pooling is a key factor for Vec2Text reconstructibility but does not necessarily contribute to improved retrieval effectiveness. We suspect that this might be due to mean pooling explicitly gathering all the token information into the embedding, whereas the CLS token embedding only implicitly learns the text information during DR training, making it less helpful for Vec2Text training.
Consequently, practitioners in the field of DR should carefully consider avoiding the use of mean pooling to enhance the security of the embedding model against Vec2Text attacks.

It is worth noting that the GTR-base model exclusively employs mean pooling and cosine similarity. This may explain why Vec2Text achieves high reconstructibility scores when trained with GTR-base embedding models.

\subsection{Zero-shot Regime and Bottleneck Pre-training}
In the previous section, we delved into fine-tuned DPR embedding models. In this section, we further explore zero-shot dense retriever with pre-trained, zero-shot embedding models. The aim is to investigate if the DR fine-tuning (done in the previous sections) itself leads to vulnerability from Vec2Text. Additionally, we explore embedding models that leverage bottleneck pre-training designed to enhance the CLS token embedding. The results are presented in Table~\ref{tab:zeroshot}.

\textit{BERT\_cls\_dot} and \textit{BERT\_mean\_cos} use the same backbone pre-trained embedding model (BERT~\cite{kenton2019bert}) and perform the exact same inference as \textit{DRP\_cls\_dot} and \textit{DRP\_mean\_cos} but without DPR training; they represent zero-shot retrievers. Their zero-shot retrieval effectiveness is significantly worse than fine-tuned DPR models (as might be expected for a zero-shot dense retriever). Mean pooling with cosine similarity exhibits higher zero-shot retrieval effectiveness than dot product with CLS token embeddings but still well below the trained DPR models from the previous section.

Now considering reconstructibility, \textit{BERT\_cls\_dot} is much less vulnerable than than \textit{DRP\_cls\_dot}.
This suggests that DR fine-tuning injects information that could aid Vec2Text in reconstruction from the CLS token embedding. In contrast, \textit{BERT\_mean\_cos} already demonstrates strong Vec2Text reconstructibility, further confirming that mean pooling with cosine similarity is potentially problematic --- even for the zero-shot setting.

We then tested the SimLM~\cite{wang-etal-2023-simlm} embedding model, which leverages a bottlenecked pre-training approach~\cite{gao-callan-2021-condenser,lu-etal-2021-less,liu2022retromae,shen2022lexmae,chuang2022diffcse,wu2022contextual,zhuang2023toroder}. Bottlenecked pre-training is a task designed to enhance the CLS token embedding. This is achieved by taking the embedding model's outputted CLS token embedding and inputting it into a weaker decoder model. The pre-training objective is to enable the decoder model to perform the MASK language modelling task or, in the case of SimLM, the ELECTRA pre-training task~\cite{clark2019electra}. Consequently, the model must learn to inject useful information into the CLS token embedding to assist the decoder in completing the task. In fact, this pre-training approach closely resembles Vec2Text training itsself, as both involve a generator that takes an embedding as input and attempts to reconstruct a piece of text. Previous studies on bottlenecked pre-training have shown improvements in DR retrieval effectiveness. However, we suspect that this pre-training approach may make the embedding particularly vulnerable to Vec2Text, raising potential privacy concerns.

Our results confirmed our hypothesis: \textit{SimLM\_cls\_dot} with additional bottlenecked pre-training on the NQ datasets\footnote{\url{https://huggingface.co/intfloat/simlm-base-wiki100w}}, demonstrated higher zero-shot retrieval effectiveness than \textit{BERT\_mean\_cos}, indicating that bottlenecked pre-training injected useful information into the CLS token that can aid in retrieval. However, the enhanced CLS token embedding proved highly vulnerable to Vec2Text. The BLEU, TF1, and COS scores were even higher than those achieved by Vec2Text trained with correct GTR-base embeddings. We conducted further tests by using SimLM as the backbone model instead of BERT to train a DPR model with the same training parameters as \textit{DPR\_cls\_dot}. The results are outlined in the row labeled \textit{SimLM\_cls\_dot\_finetuned}. Indeed, the use of SimLM further improved DPR retrieval effectiveness and also achieved very high reconstructibility. We conclude that practitioners should carefully consider the privacy implications of bottleneck pre-training before incorporating this approach.

\begin{table*}\centering
	\caption{Impact of embedding dimension and quantization. Both dimension reduction and product quantization prove to be simple and effective methods to mitigate Vec2Text reconstruction, without comprising retrieval effectiveness. There is no statistically significant difference observed in terms of retrieval effectiveness.}\label{tab:dim}
		\addtolength{\tabcolsep}{-0.2em}
%	\resizebox{1\textwidth}{!}{
		\begin{tabular}{@{}lrlrrrrrrrrr@{}}\toprule
			\multicolumn{3}{c}{\bf Embedding Models}& \bf Index size &\multicolumn{4}{c}{\bf Retrieval effectiveness} &\multicolumn{4}{c}{\bf Reconstructibility} \\
			\cmidrule(r){1-3}\cmidrule(lr){4-4}\cmidrule(lr){5-8}\cmidrule(l){9-12}
			Label & Dim. & Quantization & & top 10 &top 20 &top 100 &top 1000 &bleu &tf1 &exact &cos \\\midrule
			a. DPR\_cls\_dot (768) & 768 & None &61GB &0.748 &0.800 &0.863 &0.914 &79.2 &90.9 &43.0 &0.996 \\
			b. DPR\_cls\_dot (256) & 256 & None &21GB &0.731 &0.784 &0.855 &0.910 &28.0 &59.9 &5.9 &0.966 \\
			c. DPR\_cls\_dot (PQ\_768) & 768 & Product &16GB &0.749 &0.801 &0.862 &0.914 &2.2 &16.4 &0.0 &0.772 \\
			d. DPR\_cls\_dot (PQ\_256) & 256 & Product &5GB &0.740 &0.796 &0.864 &0.912 &2.5 &17.3 &0.0 &0.782 \\
			\bottomrule
		\end{tabular}
%	}
\end{table*}

\subsection{Embedding Dimensionality \& Quantization}
Next, we explore how embedding dimensionality and quantization, popular methods for reducing DR index size and improving retrieval efficiency, impact retrieval effectiveness and Vec2Text reconstructability.

Regarding embedding dimensionality, we trained another DPR model with the exact settings as \textit{DPR\_cls\_dot}, but adding a dense pooling layer on top to reduce the embedding dimension from 768 to 256. This variant is labeled as \textit{DPR\_cls\_dot (256)}. For embedding quantization, we applied product quantization (PQ)~\cite{jegou2010product} to \textit{DPR\_cls\_dot}. We experimented with two settings: one with the number of sub-vectors set to 768 and another with 256. The number of bits per sub-vector for both settings is set to 8 in both cases, resulting in embedding dimensions of 768 and 256. We label them \textit{DPR\_cls\_dot (PQ\_768)} and \textit{DPR\_cls\_dot (PQ\_256)} respectively.

The results presented in Table~\ref{tab:dim} indicate that dimensionality reduction and product quantization can significantly reduce the dense vector index sizes, as expected. Product quantization demonstrates robust retrieval effectiveness; when the sub-vector is set to 768, there is no decrease in top-k accuracy, and only a slight decrease in retrieval effectiveness when further reducing the dimensionality to 256. These findings align with previous studies~\cite{ma-etal-2021-simple}.

On the other hand, simply reducing the embedding dimension by adding a dense pooling layer can considerably decrease Vec2Text reconstructibility, albeit at the cost of hurting retrieval effectiveness. PQ once again demonstrates superior privacy protection ability. Both settings of PQ we considered completely mitigate Vec2Text reconstructibility. Notably, this is achieved without compromising retrieval effectiveness, and the index size is significantly reduced. Hence, we conclude that product quantization is a simple and effective method for protecting embeddings from attacks by Vec2Text.

\begin{figure}[b!]
  \includegraphics[width=.8\columnwidth]{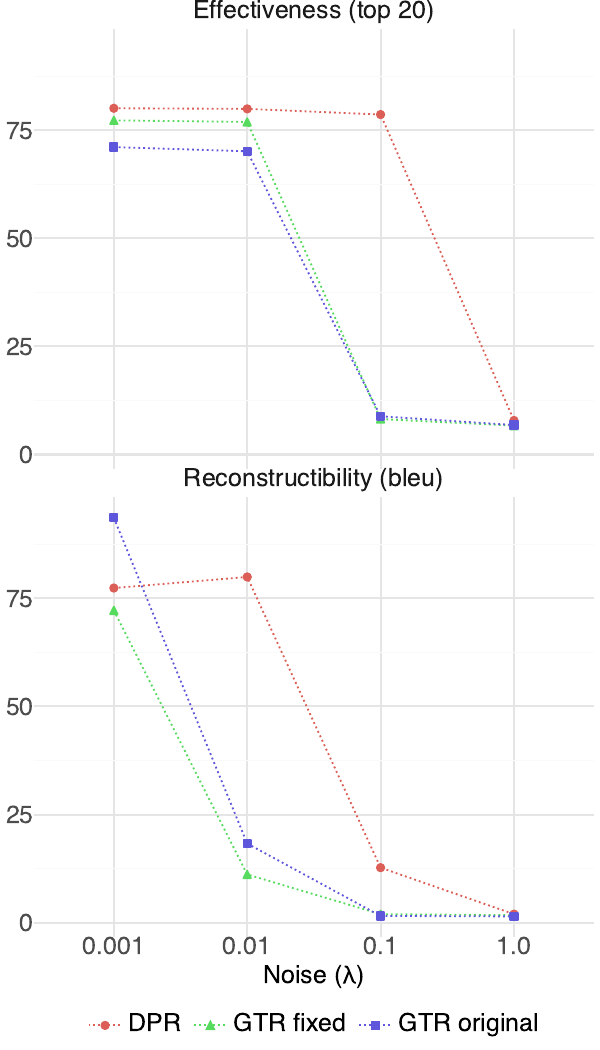}
  \caption{Impact on retrieval effectiveness and reconstructibility of different amounts of noise injection with different retrieval models. Larger $\lambda$ signifies more noise injection.}
  	\label{fig:noise_plot}
\end{figure}

\begin{table*}[!htp]\centering
	\caption{Impact of adding noise and customized embedding transformation. Noise injection was an effective reconstruction mitigation strategy. The proposed vector transformation mitigation strategy is easily applied and guaranteed to maintain retrieval effective while completely degrading reconstructibility. There is no statistically significant difference observed in terms of retrieval effectiveness.}\label{tab:tranformation}
%	\resizebox{1\textwidth}{!}{
		\begin{tabular}{@{}llrrrrrrrr@{}}\toprule
			\multicolumn{2}{c}{\bf Embedding Models}&\multicolumn{4}{c}{\bf Retrieval effectiveness} &\multicolumn{4}{c}{\bf Reconstructibility} \\
			\cmidrule(r){1-2} \cmidrule(lr){3-6}\cmidrule(l){7-10}
			Label & Mitigation & top 10 &top 20 &top 100 &top 1000 &bleu &tf1 &exact &cos \\\midrule
			a. DPR\_cls\_dot & None &0.748 &0.800 &0.863 &0.914 &85.3 &93.8 &54.6 &0.998 \\
			b. DPR\_cls\_dot\_noise & Noise injection &0.730 &0.790 &0.856 &0.911 &12.7 &45.5 &0.0 &0.879 \\
			c. DPR\_cls\_dot\_transform & Vector transformation &0.748 &0.800 &0.863 &0.914 &1.1 &6.0 &0.0 &0.521 \\
			\bottomrule
		\end{tabular}
%	}
\end{table*}

\begin{table*}[!htp]\centering
	\caption{Results of adding noise during Text2Text training. The target embedding model is $DPR\_cls\_dot$. Noise injection proved to be an effective mitigation strategy even when Vec2Text is actually trained with noisy embeddings.}\label{tab:noise_train}
		\begin{tabular}{@{}llrrrrrrrr@{}}\toprule
			\multicolumn{2}{@{}c}{\bf Vec2Text Models} & \multicolumn{8}{c}{\bf Reconstructibility}\\
			& & \multicolumn{4}{c}{On original embeddings} & \multicolumn{4}{c}{On noisy embeddings} \\ 
			\cmidrule(r){1-2} \cmidrule(lr){3-6}\cmidrule(l){7-10}
			Label & Trained on &bleu &tf1 &exact &cos &bleu &tf1 &exact &cos \\\midrule
			Vec2Text & Original embeddings &85.3 &93.8 &54.6 &0.998 &12.7 &45.5 &0.0 &0.879 \\
			Vec2Text\_noise & Noisy embeddings &27.2 &62.6 &1.3 &0.979 &21.0 &55.3&0.7 &0.908 \\
			\bottomrule
		\end{tabular}
\end{table*}

\section{Mitigation Strategies}
In the previous sections, we explored how different DR training and inference strategies trade-off retrieval effectiveness and embedding privacy protection. In this section, we evaluate methods that proactively protect against the Vec2Text attacks.

\subsection{Noise Injection}
In the original Vec2Text paper, \citeauthor{morris-etal-2023-text} explored a basic mitigation strategy of adding Gaussian noise directly to each embedding:
\begin{equation}
	\phi_{noisy}(x) = \phi(x) + \lambda \cdot \epsilon, \epsilon \sim  \mathcal{N}(0, 1),
\end{equation}
where $\phi(x)$ is the embedding for a given input text $x$, $\epsilon$ is the noise vector whose elements are sampled from a Gaussian distribution, and $\lambda$ is a hyperparameter controlling the amount of injected noise.  \citeauthor{morris-etal-2023-text} showed that setting $\lambda = 0.1$ completely prevents reconstructibility for the GTR-base embedding model, but it also has a major negative impact on retrieval effectiveness.

%would destroy both retrieval effectiveness and Vec2Text reconstructibility for the GTR-base embedding model (with incorrect embeddings).

Here, we reproduce the noise injection mitigation strategy on our trained $DPR\-\_cls\_dot$ embedding model, labeled as $DPR\_cls\_dot\_noise$ in Table~\ref{tab:tranformation}. Our results show partially different outcomes: setting $\lambda = 0.1$ did indeed prevent Vec2Text reconstructibility but only slightly impacted retrieval effectiveness. 
This may be due to the sensitivity of the specific embedding models to the amount of noise (as controlled by $\lambda$). To better understand this we completed a sweep of $\lambda = [0.001, 0.01, 0.1, 1.0]$ for DPR, GTR and GTR with corrected embeddings; this is shown in Figure~\ref{fig:noise_plot}. We observe that DPR requires more noise to see an effect: both in terms of retrieval effectiveness deterioration, and reconstructibility robustness. We hypothesise that this is due to the different distance metric used by the two methods (GTR uses cosine, while DPR here uses dot product).
Overall, adding noise remains a desirable mitigation strategy but correct parameter setting is critical. However, the correct setting of the noise parameter appears to be dependent on the specific embedding model used. In the next section we propose a new mitigation strategy that does not have this drawback.

A question still remains, though, of whether adding noise was an effective mitigation strategy if Vec2Text was actually trained using these noisy vectors. 
To investigate this, we re-trained a Vec2Text model while adding the same Gaussian noise to the input embeddings. The results are presented in Table~\ref{tab:noise_train}. Our findings demonstrate that adding noise to Vec2Text training reduces reconstructibility on original embeddings. While it does make Vec2Text more robust to noisy embeddings, the level of reconstructibility remains very low. Thus, we conclude that adding noise to Vec2Text training does not make the noise-adding defence vulnerable.

\subsection{Embedding Transformation}

Finally, we propose a new mitigation strategy that prevents reconstruction via Vec2Text but has the benefit of theoretically guaranteeing no change in retrieval effectiveness and not requiring parameter tuning that is dependent on the embedding model used. 

Our idea is straightforward: we propose applying a simple transformation to all the embeddings, with a transform function customized and only visible to each user (similar to an API key) who sends text to the embedding model APIs (such as OpenAI users sending text to OpenAI APIs to obtain embeddings). Formally, we define the transformed embeddings as:
\begin{equation}
	\phi_{\text{transformed}}(x) = f(\phi(x)),
\end{equation}
where $f$ is a linear transformation function unique to each user. For example, in our experiments, we simply set $ f(\phi(x)) = -2.6 \cdot \phi(x)$ so all embeddings are adjusted by a constant value. The function (in our case the constant value) is kept secret and thus not known to a Vec2Text attacker. This means that attackers will obtain different embeddings from the API calls to train their Vec2Text model. Additionally, on the user side, since the same transformation function is applied to all embeddings retrieval will not be impacted. Users can still reverse engineer to get the original embeddings since the transformation function is visible to the user.

We present our empirical results in Table~\ref{tab:tranformation}, labeled with \textit{DPR\-\_cls\_dot\_transform} (row c.). Note that the retrieval effective is identical to that of \textit{DPR\_cls\_dot} (row a.), as guaranteed by the uniform transformation. At same time, reconstructibility is almost completely degraded so privacy is protected. We believe this simple strategy could be easily adopted by service providers like OpenAI or Cohere to offer their users the option to make their embeddings more secure.

\section{Conclusion}
%\todo{Future works: how Vec2Text response to sparse embeddings? How to protect corpus poisoning attack from Vec2Text. }

Dense retrievers have proven an effective and efficient retrieval method and are now widely adopted in working systems. Much of the benefit comes from using text embeddings to represent and compare information. However, the reliance on text embeddings also opens up dense retrievers to possible threats that exploit such embeddings. Methods like Vec2Text, which can successfully reconstruct the original text from an embedding, could pose serious privacy risks, especially now embeddings are made publicly available via APIs (e.g., OpenAI or Cohere).

%This paper reproduces and extends Vec2Text. We begin by uncovering an issue in the original Vec2Text implementation whereby the last hidden layer rather than the dense pooling layer was used for the final text embeddings for one of the considered embedding models. We correct this issue, rerunning the original experiments and showing that Vec2Text is even more effective in retrieval and reconstructibility.

This paper reproduces and extends Vec2Text. We begin by uncovering two issues in the original Vec2Text experimental setting. One issue is related to the Vec2Text implementation, whereby the last hidden layer rather than the dense pooling layer was used for the final text embeddings for one of the considered embedding models. The other issue is a serious data leakage problem, where more than 50\% of validation data points also appear in the Vec2Text training data. We correct these issues and re-executed the original experiments, providing a more accurate evaluation.

We then apply Vec2text under a number of new conditions: different distance metrics, pooling functions, bottleneck pre-training, embeddings size and model quantization. We discover that the two techniques of mean pooling and bottleneck pre-training increases the risk of reconstructibility. We also show that even zero-shot rankers with poor retrieval effectiveness are vulnerable to Vec2Text reconstruction. A surprising finding was that the compression techniques of embeddings dimension reduction and model product quantization prove to be effective at significantly reducing reconstructibility, while maintaining retrieval effectiveness. Overall, our results give a comprehensive insight into the trade-offs between privacy protection and retrieval effectiveness.

Having clearly outlined the risk, we explore mitigation strategies. We reproduce an existing mitigation strategy that adds noise to the text embeddings, while extending its test to verify if Vec2Text could circumvent this mitigation strategy by training with these noisy embeddings. We show noise injection is a valid mitigation strategy, though comes with trade-offs in terms of retrieval effectiveness and sensitivity to parameter choices. We then propose a new embedding transformation mitigation strategy guaranteed to maintain retrieval effective while completely degrading reconstructibility so privacy is protected. We believe this simple strategy could be easily adopted by services like OpenAI or Cohere to provide their users with an option to make their embeddings more secure.

Overall, this study maps out under what specific conditions Vec2Text could pose a potential threat to current dense retrieval systems, while also highlighting different mitigation strategies and the trade-off between retrieval effectiveness and reconstructibility. While Vec2Text may be a threat, there are some effective and easy to integrate methods to patch current dense retrievers.

%\subsection*{Acknowledgements} We thank \citeauthor{morris-etal-2023-text} for giving permission to re-use their figure illustrating the Vec2Text method, and for extensive assistance in clarifying the original codebase. We are also thankful to [Anonymised] from [Anonymised] who examined an earlier version of our GitHub repository and spotted potential issues that helped us identifying the data leakage problem in the original work. 
%Our work is partially funding by \texttt{Anonymized}.

\begin{acks}
We thank Morris et al. for giving permission to reuse their figure illustrating the Vec2Text method. We also thank Xin Cheng from Peking University for reviewing our code and identifying the data leakage issue in the original NQ dataset used for training Vec2Text models.
This research is partially funded by Beijing Baidu Netcom Technology Co, Ltd, for the project "Federated Online Learning of Neural Rankers", under funding schema 2022 CCF-Baidu Pinecone.
\end{acks}

%\clearpage
\balance
\bibliographystyle{ACM-Reference-Format}
\bibliography{references}

\appendix

\end{document}